\documentclass[sigconf]{acmart}

\usepackage{microtype}
\usepackage{graphicx}
\usepackage{booktabs} 
\usepackage{xspace}
\usepackage{graphicx} 
\usepackage{xspace}
\usepackage{xcolor}
\usepackage{mfirstuc}
\usepackage{enumitem}
\usepackage{color}
\usepackage{tikz}
\usepackage{comment}
\usepackage{hyperref}
\usepackage{multirow}
\usepackage{listings}
\usepackage{array}
\usepackage{booktabs}
\usepackage{placeins}
\usepackage{titlesec}
\titlespacing*{\section}{0pt}{0.7ex}{0.7ex}
\titlespacing*{\subsection}{0pt}{0.6ex}{0.6ex}
\titlespacing*{\subsubsection}{0pt}{0.5ex}{0.5ex}

\usepackage{hyperref}
\usepackage{subcaption}
\usepackage{algorithm}

\usepackage{algpseudocode}
\usepackage{hyperref}
\usepackage{url}
\usepackage{cleveref}

\renewcommand\footnotetextcopyrightpermission[1]{} 
\AtBeginDocument{%
  }

\setcopyright{acmlicensed}
\copyrightyear{2018}
\acmYear{2018}
\acmDOI{XXXXXXX.XXXXXXX}

\acmConference[Conference acronym 'XX]{Make sure to enter the correct
  conference title from your rights confirmation emai}{June 03--05,
  2018}{Woodstock, NY}
\acmISBN{978-1-4503-XXXX-X/18/06}

\newcommand{\meta}{Meta\xspace}
\newcommand{\mthreen}{Lattice\xspace}
\newcommand{\metalattice}{\meta \mthreen}
\newcommand{\sumabbr}{\mthreen Network\xspace}
\newcommand{\sumabbrs}{\mthreen Networks\xspace}

\newcommand{\kdap}{\mthreen KTAP\xspace}

\newcommand{\mdmo}{MDMO\xspace}
\newcommand{\mpartition}{\mthreen Partitioner\xspace}

\newcommand{\maw}{\mthreen Zipper\xspace}

\newcommand{\pofs}{\mthreen Filter\xspace}
\newcommand{\pofsnomthreen}{Filter\xspace}

\newcommand{\metasketch}{\mthreen Sketch\xspace}

\newcommand{\circled}[1]{\raisebox{.5pt}{\textcircled{\raisebox{-.9pt}{#1}}}}

\title{\textsc{\metalattice}: Model Space Redesign for Cost-Effective Industry-Scale Ads Recommendations}

\author{Liang Luo, Yuxin Chen, Zhengyu Zhang, Mengyue Hang, Andrew Gu, Buyun Zhang, Boyang Liu, Chen Chen, Fan Yang, Feifan Gu, Huayu Li, Jade Nie, Jiayi Xu, Jiyan Yang, Jongsoo Park, Laming Chen, Longhao Jin, Qin Huang, Shali Jiang, Shiwen Shen, Shuaiwen Wang, Siyang Yuan, Tongyi Tang, Weilin Zhang, Xi Liu, Xiaohan Wei, Yuchen Hao, Xiaozhen Xia, Yasmine Badr, Zeliang Chen, Chengze Fan, Dong Liang, Qianru Li, Sihan Zeng, Wenjun Wang, Yunlong He, Yinbin Ma, Maxim Naumov, Yantao Yao, Wenlin Chen, Santanu Kokay, GP Musumeci, Ellie Wen}
\affiliation{%
  \institution{Meta AI}
  \city{Menlo Park}
  \state{California}
  \country{USA}
}
\renewcommand{\shortauthors}{Meta Lattice Team}

\begin{abstract}
The rapidly evolving landscape of products, surfaces, policies, and regulations poses significant challenges for deploying state-of-the-art recommendation models at industry scale, primarily due to data fragmentation across domains and escalating infrastructure costs that hinder sustained quality improvements.

To address this challenge, we propose \mthreen, a recommendation framework centered around \textit{model space redesign} that extends Multi-Domain, Multi-Objective (\mdmo) learning beyond models and learning objectives. \mthreen addresses these challenges through a comprehensive model space redesign that combines cross-domain knowledge sharing, data consolidation, model unification, distillation, and system optimizations to achieve significant improvements in both quality and cost-efficiency.

Our deployment of \mthreen at \meta has resulted in 10\% revenue-driving top-line metrics gain, 11.5\% user satisfaction improvement, 6\% boost in conversion rate, with 20\% capacity saving. 
\end{abstract}

\begin{document}




\maketitle

\begin{sloppypar}
\section{Introduction}

\begin{figure*}[t]
    \begin{center}
    \includegraphics[width=0.93\linewidth]{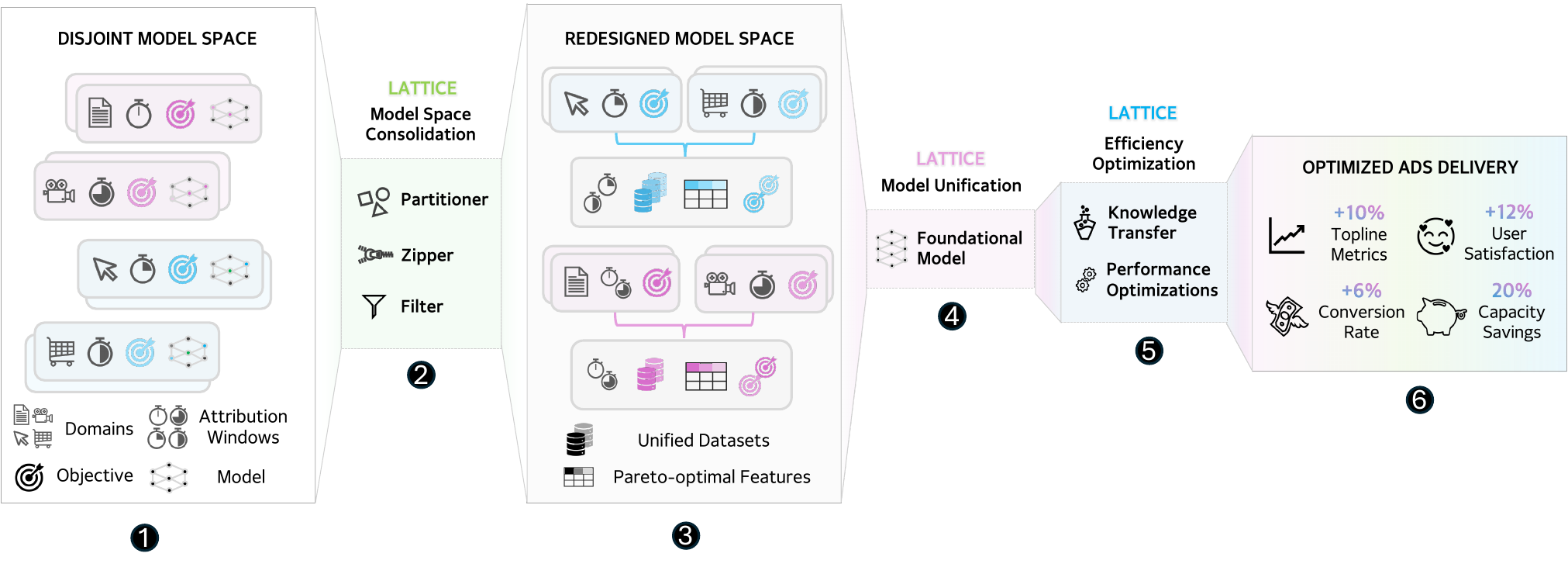}
    \vspace{-2em}
    \caption{\mthreen overview: \circled{1} Even with \mdmo frameworks, current model space remains scattered, with heterogeneous datasets, assorted attribution windows and siloed models and objectives. \circled{2}-\circled{3} \mthreen redesigns the model space by consolidating datasets of heterogeneous formats, attribution windows and selects features on the pareto-front to form a small set of consolidated portfolios, improving knowledge sharing across domains and cutting down serving cost. \circled{4} Foundational \sumabbr employs a \mdmo-based, unified model architecture to consume multiple data formats and produces predictions for all consolidated objectives. \circled{5} These models are further distilled to lean user-facing models via novel knowledge transfer and efficiency optimizations. \circled{6} Deployed \mthreen at Meta results in significant improvements in all key metrics and  cost reduction.}
    \vspace{-1em}    
    \label{fig:m3n_overview}        
    \end{center}
\end{figure*}

The discovery of scaling laws in recommendation models~\cite{zhang2024wukong,ardalani2022understanding,geng2023vip5,guo2023embedding} has unveiled new avenues for enhancing recommendation performance. However, translating these theoretical insights into practical deployment at industry scale faces challenges.

The first challenge is economic scalability. Modern recommenders operate across thousands of domain-objective pairs (called \textit{portfolios}), each traditionally requiring a dedicated model~\citep{zhang2024wukong, dlrm, wang2021dcn, afn, mao2023finalmlp}. While scaling laws demonstrate clear performance benefits from larger models~\cite{zhang2024wukong, kaplan2020scaling}, the prohibitive infrastructure costs make it impractical to upscale every model independently. This forces practitioners to focus optimization on only a small subset of key portfolios, leaving significant performance gains unrealized.

The second challenge is data fragmentation. Large-scale models are inherently data-hungry, yet maintaining separate datasets for each portfolio creates artificial barriers to knowledge sharing. This fragmentation restricts cross-product understanding of user preferences and limits the data available to individual models, creating a scalability bottleneck due to data scarcity in specialized domains.

The third challenge is deployment constraints. Industrial production environments impose stringent inference latency requirements that severely limit the complexity of models that can be practically deployed, creating a fundamental tension between model capability and operational feasibility.

To address these interconnected challenges and unlock the full potential of large recommendation models, we propose moving beyond traditional per-portfolio scaling toward model space redesign via \mthreen (Figure~\ref{fig:m3n_overview})—a transformative framework that tackles the three fundamental challenges.

For economic scalability, \mthreen reduces the overall number of models required by consolidating portfolios to harness scaling laws while minimizing costs—a single invocation on \mthreen models can simultaneously generate predictions for multiple portfolios. For data fragmentation, \mthreen enriches training data per model by facilitating cross-portfolio sharing, allowing models to learn from broader and more diverse datasets. For deployment constraints, \mthreen overcomes latency barriers through a hierarchical model space design where large-scale foundational models~\citep{exfm} are decoupled from real-time serving and act as teachers, transferring knowledge to smaller, highly-optimized user-facing models. 

However, effective portfolio consolidation introduces technical challenges that extend far beyond traditional \mdmo approaches, which primarily focus on tweaking model architectures and learning objectives while assuming clean, readily available datasets. 

First, the delayed feedback problem: Ad recommenders face different attribution windows (time windows for collecting positive signals~\citep{chen2022asymptotically, ktena2019addressing, wang2020delayed}), creating a critical trade-off between data freshness and correctness. \mthreen addresses this through \textit{\maw}, which integrates datasets by associating each ad impression with a randomly selected attribution window, infusing both freshness and correctness into unified datasets.

Second, the feature selection problem: Ensuring model performance across consolidated portfolios while adhering to resource constraints requires optimal feature selection from vastly expanded feature spaces. \mthreen employs \textit{\pofs}, a Pareto-optimal algorithm that selects features from merged datasets to guarantee pareto-optimal quality across all portfolios.

Third, the architecture design problem: With data and features consolidated, \mthreen constructs \textit{\sumabbrs}, a family of \mdmo models featuring novel architectures tailored to handle diverse input formats through interleaved learning while mitigating domain conflicts via parameter untying.

Fourth, the knowledge transfer problem: To fully harness scaling laws without sacrificing inference performance, \mthreen implements \textit{\kdap}, a highly-scalable mechanism leveraging asynchronous precompute for hierarchical knowledge transfer at inference time, augmenting knowledge distillation efficacy while maintaining real-time serving requirements.

The unified model architecture enables focused efficiency optimizations. We develop \textit{\metasketch}, an automated search tool for model hyperparameters and parallelization strategies, guided by scaling laws to balance quality and performance. Additionally, the consolidated model space requires only a small number of models, enabling careful optimization through customized GPU kernels and low-precision training and inference.

Finally, \mthreen incorporates \textit{\mpartition}, a policy tool that 
guides portfolio consolidation, improving knowledge sharing, reducing data conflicts, and maximizing global model performance within resource budgets while ensuring privacy compliance.

We evaluate \mthreen through extensive experiments across both public benchmarks and industry-scale datasets with real-world recommendation scenarios. Our results demonstrate that \mthreen consistently outperforms 10 state-of-the-art baselines, achieving up to 1\% improvement in prediction loss while delivering up to 1.3$\times$ hardware efficiency gains on 1024 GPUs. Our production deployment of \mthreen across a representative set of Meta's ads model types has delivered substantial real-world impact: 10\% improvement in revenue-driving top-line metrics, 11.5\% uplift in user satisfaction, 6\% boost in conversion rate, accompanied by 20\% capacity savings, demonstrating that \mthreen successfully bridges the gap between theoretical advances and practical industry deployment at scale.
\section{Related Work and Opportunities}
\label{sec:background}
This section provides a review on the recent advancements and further opportunities to improve with \mthreen.

\subsection{Portfolio Consolidation} 

\textbf{Status Quo} 
Current recommendation systems optimize narrow objective sets within specific domains~\citep{zhang2024wukong, dlrm, wang2021dcn, afn, mao2023finalmlp}, becoming inefficient as domains and objectives proliferate.

Multi-Domain, Multi-Objective (\mdmo) approaches~\citep{li2020ddtcdr, yan2019deepapf, ma2018modeling, sheng2021one, wang2022causalint, yang2022adasparse, tang2020progressive, liu2022multi, malhotra2022dropped, li2023adatt, yang2023adatask, wang2021understanding, wang2022can} use models consuming all-domain data for all tasks, but monolithic designs limit model complexity due to latency constraints and task interference issues~\citep{tang2023improving, he2022metabalance}. Traditional ensemble approaches that combine separate models increase rather than reduce computational costs.

\noindent\textbf{Opportunities} Portfolio consolidation extends \mdmo to broader settings, capturing massive domains without compromising deployability through multiple foundational models within latency envelopes while separating competing domains and objectives to mitigate interference~\citep{he2022metabalance, tang2023improving}.




\subsection{Data Integration} 
\noindent\textbf{Status Quo} Most \mdmo research assumes clean, readily available datasets~\citep{10.1145/3488560.3498479, 10.1145/3580305.3599884}, simulating \mdmo setups by partitioning coherent datasets~\citep{zhang2024m3oe, 10.1145/3604915.3608828}. Real-world datasets are heterogeneous (diverse data types and formats~\citep{zhai2024actions}), fragmented across product surfaces, sparse due to rare positive signals, and dynamic with shifting user behaviors. For example, delayed feedback creates additional challenges where different attribution windows introduce trade-offs between label completeness and freshness~\citep{chen2022asymptotically, ktena2019addressing, wang2020delayed}.

\noindent\textbf{Opportunities} Data integration enables sustainable model scaling by addressing data scarcity while enhancing model quality through cross-domain knowledge sharing, optimal feature selection, and effective attribution window blending.

\subsection{Model Unification} 

\noindent\textbf{Status Quo} Representative \mdmo models use a common-then-specialize paradigm, including MMoE~\citep{ma2018modeling} with multi-gate mixture-of-experts, M2M~\citep{10.1145/3488560.3498479} using meta units and tower modules, M$^3$oE~\citep{zhang2024m3oe} with two-level feature extraction mechanisms, PEPNet~\citep{10.1145/3580305.3599884} employing personalized embedding networks, and M3REC~\citep{10.1145/3604915.3608828} using meta-learning for unified embedding representation. However, these approaches assume homogeneous input formats and struggle with cross-domain interference~\citep{tang2023improving}. 

\noindent\textbf{Opportunities} Portfolio consolidation requires architectures that efficiently process heterogeneous input formats—dense, sparse, and sequential data~\citep{zhai2024actions, liu2024multimodalrecommendersystemssurvey}—while mitigating learning interference across consolidated domains through novel architectural designs and parameter isolation techniques.

\subsection{Efficiency Improvement}

\noindent\textbf{Status Quo} Recommendation models face efficiency challenges from large embedding tables~\citep{persia} and poor hardware utilization~\citep{dmt}. Existing approaches target architecture~\citep{zhang2024wukong, song2019autoint, liu2024mamba4rec}, topology-aware designs~\citep{dmt, neuroshard, zhang2022dhen}, specialized hardware~\citep{Ournextg17:online}, and knowledge distillation~\citep{hinton2015distilling}. Training stability from distribution shifts and \mdmo learning are addressed through gradient clipping~\citep{pascanu2013difficulty, tang2023improving}, better feature interactions~\citep{adnan2023adrec}, and normalization~\citep{ba2016layernormalization, santurkar2018does}. 

\noindent\textbf{Opportunities} Portfolio consolidation enables focused efficiency optimizations across fewer, unified models while requiring joint optimization strategies that traditional approaches cannot provide, for example, low-precision training and inference, successful in LLMs~\citep{deepseekai2025deepseekv3technicalreport}, remain largely unexplored in recommendation due to scattered model landscape as they need to be tuned per-model.

\begin{figure*}[t]
    \begin{center}
       \includegraphics[width=.83\textwidth]{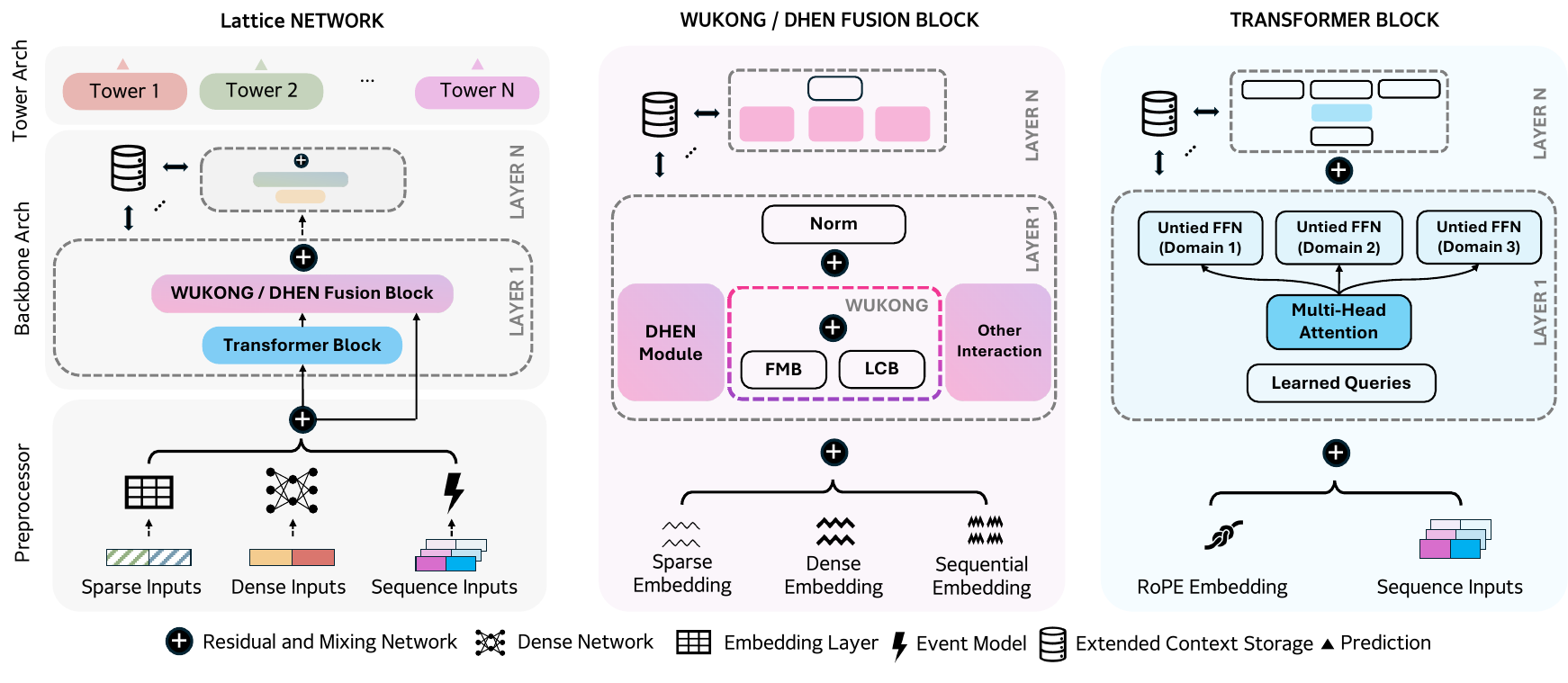}
    \vspace{-1.5em}       
    \caption{\sumabbrs interleave sequence and non-sequence learning to handle diverse input types in merged domains.}
    \vspace{-1.5em}    
    \label{fig:afm_model_arch}       
    \end{center}
\end{figure*}

\section{\metalattice}
We present \mthreen, a comprehensive framework that advances state-of-the-art recommendation at industry scale through co-design of data, model, and efficiency strategies.

\subsection{Portfolio Consolidation: \mpartition}
\label{sec:partitioner}

Grounded in established theories that sample complexity and estimation error decrease when tasks are related~\cite{Baxter_2000, 10.5555/1248547.1248552}, we partition domain-objective pairs into manageable recommendation groups, reducing the total number of required models while leveraging foundational model~\cite{bommasani2021opportunities}, auxiliary learning~\citep{he2022metabalance}, multi-task learning~\citep{li2020ddtcdr}, and transfer learning~\citep{5288526} theories to address data fragmentation and improve cross-domain knowledge sharing~\citep{ma2018entire,mordan2018revisiting}.

When grouped, domain-objective pairs merge their associated datasets, features, and objectives, requiring only a single model per group. However, naive consolidation causes training instability and quality degradation due to domain conflicts~\citep{tang2023clippy,he2022metabalance}. 

\mpartition addresses this through a policy-driven approach that:

\begin{itemize}[leftmargin=*,noitemsep,topsep=0pt]
  \item \textbf{Prioritizes overlapping ID spaces:} Merges domains with substantial user/item overlap (e.g., combining "News Ads" and "Video Ads" which share users, versus keeping "Marketplace" separate due to distinct item) to facilitate knowledge sharing.
  \item \textbf{Groups objectives by feedback characteristics:} Separates (1) fresh, dense feedback tasks (click, like, follow) from (2) delayed, sparse feedback tasks (purchase, conversion). Task similarity within groups is assessed via empirical loss weighting~\cite{pentina2017multi,murugesan2016adaptive} or gradient-based metrics~\cite{achille2019task2vec, fifty2021efficiently}, and subgroups can be further created if similarity is low. Conflicts within a group are mitigated through parameter untying (\S\ref{sec:arch-change}), supervision changes (\S\ref{sec:supervision-change}), and MetaBalance~\cite{he2022metabalance}.
  \item \textbf{Ensures compliance:} Respects privacy policies and data sharing restrictions.
  \item \textbf{Allocates resources:} Distributes compute and storage budgets based on estimated revenue impact.
\end{itemize}
\subsection{Data Integration: \maw \& \pofsnomthreen}
\label{sec:data_consolidation}
\mthreen creates unified datasets and selects relevant features to facilitate cross-domain knowledge sharing and mitigate data scarcity for consolidated recommendation portfolios.

\noindent\textbf{Unified Dataset Construction:} We concatenate multiple datasets, where the feature set becomes the union of all features. Each record contains an original entry from one domain with non-existent features zero-padded, enabling models to learn cross-domain patterns while handling heterogeneous feature spaces.

\subsubsection{\textbf{Mixing Attribution Windows via \maw}}

Unlike typical recommenders that focus on real-time user feedback, ad recommenders face delayed feedback challenges. User conversions (e.g., purchases) can occur anywhere from minutes to days after an ad impression. To capture these delayed conversions, advertisers define \textit{attribution windows}—time periods during which post-impression actions are credited to the ad. However, this creates a fundamental trade-off: shorter windows provide fresher training data but miss delayed conversions, while longer windows capture more conversions but introduce staleness.

Traditionally, this is addressed by ensembling $K$ models trained on datasets with different attribution windows, or through multi-pass training that updates models as attribution windows close. Both increase training costs by $K\times$ and can cause overfitting or instability due to conflicting labels for the same impression~\cite{kuaishou-multipass}.

Instead of maintaining $K$ separate datasets, \maw creates a single unified dataset by randomly assigning each impression to one attribution window based on a tunable probability distribution (usually uniform random). The assignment uses deterministic hashing of the impression signature (user ID, ad ID, timestamp). 

We then modify the model architecture to include separate prediction heads for each attribution window incorporated in the dataset. During training, impressions are routed to their assigned prediction head, allowing the model to learn from data at different freshness-correctness trade-off points simultaneously. The longest attribution window serves as the "oracle" head, learning from the most complete data, while shorter windows provide fresher signals that improve the shared backbone representation.

At serving time, we use only the oracle prediction head, which benefits from both the complete long-window data and the fresher shorter windows during training.

\subsubsection{\textbf{Pareto-Optimal Feature Selection via \pofs}}

Although tens of thousands of features characterizing users and items are available, resource constraints limit models to using only selective features (typically thousands). Portfolio consolidation introduces a new challenge: selecting features that achieve optimal performance across \textit{multiple} consolidated portfolios simultaneously, where each portfolio may value different feature subsets.

Standard feature selection approaches optimize for single tasks or use weighted combinations across tasks, which can result in degradation of specific portfolios when one dominates the optimization objective. This is problematic when consolidated portfolios have different business importance or data characteristics.

\pofs addresses this through Pareto-optimal selection, ensuring no portfolio is unfairly penalized while maximizing overall quality across all consolidated portfolios.

\pofs begins by computing a feature importance score vector for each feature in the set $\mathcal{F}$. For the $i$-th feature across $N$ tasks (representing all objectives from consolidated portfolios), we denote its importance scores as $\mathbf{F}_i = (f_{i,1}, \ldots, f_{i,N})$, where $f_{i,j}$ represents the importance of feature $i$ in task $j$, computed using permutation-based importance~\citep{breiman2001random}.

We establish a partial order relationship "dominated by" ($\preccurlyeq$): $\mathbf{F}_i \preccurlyeq \mathbf{F}_k$ if and only if $f_{i,j} \leq f_{k,j}$ for all $j \in [1, N]$. When feature $k$'s importance vector dominates feature $i$, feature $i$ is not on the current pareto front and hence excluded from immediate selection.

\begin{figure}
    \centering
    \includegraphics[width=0.8\columnwidth]{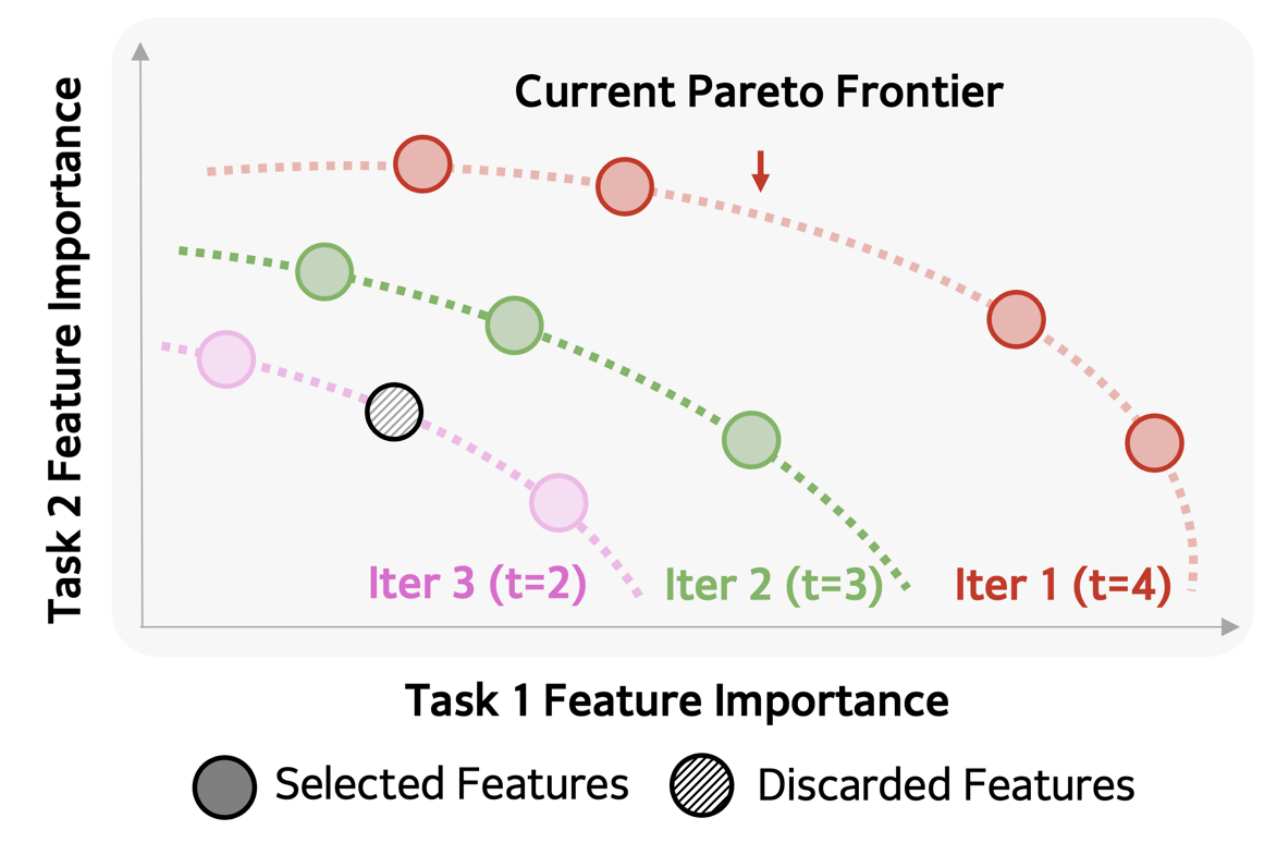}
    \caption{Illustration: \pofs with $T=9$ and $|\mathcal{F}|=10$ via 3 iterations.}
    \vspace{-2.4em}
    \label{fig:cherrypick}  
\end{figure}

Given target feature count $T$, \pofs iteratively identifies features on the current Pareto frontier. In each iteration, features from the current frontier are selected. If there are more features on the pareto frontier than the budget, remaining quota is filled by randomly picking features on the current frontier. This random process ensures quality because features are pre-sorted by importance, critical ones are chosen in earlier iterations, they should have already appeared on previous frontiers. It also does not introduce bias because selection freezes the current frontier. This process continues until $T$ features are selected. \Cref{fig:cherrypick} provides an illustration of \pofs.

\pofs ensures balanced performance across all portfolios. The approach can be extended to incorporate additional criteria such as feature computation cost, storage requirements, portfolio importance, and data freshness, enabling rich multi-objective optimization for feature selection.
\subsection{Model Unification: \sumabbrs}
\label{sec:model}

Each \sumabbr covers one recommendation group. When domains with different input formats are merged, a unified model architecture must handle diverse input types while maintaining the ability to learn domain-specific patterns. 

\sumabbrs process mix-format, multi-modal inputs: categorical features ($\mathcal{F}_c$) such as user/item IDs, dense features ($\mathcal{F}_d$) like user age or item price, and sequence features ($\mathcal{F}_s$) such as user interaction histories or raw contents. Each sample in a batch $B$ produces one prediction per consolidated task.

\sumabbrs adopt a novel three-stage preprocessor-backbone-task architecture with interleaving of sequence and non-sequence processing plus parameter untying (Figure~\ref{fig:afm_model_arch}). This design first unifies disparate input representations, then performs cross-domain interaction through specialized modules, and finally generates task-specific predictions.

\subsubsection{\textbf{Feature Processors}}
unify input representations and project them into a common embedding space with uniform dimension $d$, enabling subsequent modules to operate on standardized inputs for all original data types:

\begin{itemize}[leftmargin=*,noitemsep,topsep=0pt]
  \item \textbf{Categorical Features ($\mathcal{F}_c$):} Processed through embedding tables, producing output $O_c$ with shape $(B, |\mathcal{F}_c|, d)$.

  \item \textbf{Dense Features ($\mathcal{F}_d$):} Processed by MLPs to handle numerical inputs, producing output $O_d$ with shape $(B, |\mathcal{F}_d|, d)$.

  \item \textbf{Sequence Features ($\mathcal{F}_s$):} Processed by attention-based event models. This produces sequence embeddings $O_s$ with shape $(B, |\mathcal{F}_s| \times K, d)$, where $K$ represents the output feature count per event source.
\end{itemize}

A mixing network then concatenates $O_c$ and $O_d$ and normalizes them to form unified non-sequence representation $O_{cd}$, while preserving $O_s$ separately. This separation enables the backbone to apply interleaved learning. 

\subsubsection{\textbf{Backbone}} 
\label{sec:network_backbone}
The \sumabbr backbone is designed for efficient dense scaling~\citep{zhang2024wukong, shin2023scaling, zhang2023scaling}, leveraging modern hardware's superior compute capabilities over memory and network resources~\citep{luo2018parameter, luo2020plink}. The architecture addresses the core challenge of processing both sequence and non-sequence data effectively through interleaved learning.

\noindent{\textbf{Extended Context Storage (ECS)}} provides a global key-value store supporting DenseNet~\citep{8099726}-style residual connections and intermediate activation access, enabling high-bandwidth information flow across layers and components which significantly helps with deeper networks~\citep{xiao2025muddformerbreakingresidualbottlenecks}.

\noindent{\textbf{Transformer Blocks (TB)}} process RoPE-encoded~\citep{su2023roformerenhancedtransformerrotary} input sequences $O_s$, producing contextualized sequences $O_s'$ using standard transformer layers~\citep{vaswani2017attention}: $O_s' = TB(ROPE(O_s))$. To enable better user-ad feature interactions, cross-attention layers and adaptive parameter generation networks~\citep{NEURIPS2022_9cd0c571} are adopted in FFN layers.

\noindent{\textbf{DHEN/Wukong Fusion Blocks (DWFB)}} address transformer limitations in processing non-sequence data by capturing bit-wise interactions~\citep{wang2021dcn}. DWFBs flatten $O_s'$, concatenate with $O_{cd}$, and produce updated non-sequence representations $O_{cd}'$ using Factorization Machine Blocks (FMB) and Linear Compression Blocks (LCB)~\citep{zhang2022dhen, zhang2024wukong}: $O_{cd}' = DWFB([O_s';O_{cd}])$.

Evidently, each backbone block applies specialized processing to sequence versus non-sequence data by different modules that are optimized for different data modalities, facilitating the higher-level interactions between sequences and non-sequences~\citep{interformer}.

\subsubsection{\textbf{Task Modules}}
\label{sec:tower_arch}
Task-specific adaptation uses lightweight MLP layers (one per objective) that project shared backbone embeddings to task-specific output spaces, enabling specialization while maintaining shared representations.

\subsubsection{\textbf{Architecture Changes}}
\label{sec:arch-change}

To address instability issues inherent in \mdmo training across diverse domains and objectives, we employ several architectural techniques:

\noindent\textbf{Modality Contention Mitigation}: On mixed-modal datasets, we mitigate contention among different modalities~\citep{chameleonteam2024chameleon} via QK-norm~\citep{dehghani2023scaling} in attention mechanisms.

\noindent\textbf{Parameter Untying for Domain Conflicts}: \sumabbrs adopt parameter untying~\citep{lin2024momaefficientearlyfusionpretraining,liang2024mixtureoftransformerssparsescalablearchitecture,deepseek_janus,xin2025i2moeinterpretablemultimodalinteractionaware,liao2025mogaoomnifoundationmodel,deng2025bagel} to mitigate learning conflicts not resolved by \mpartition. By dedicating weights for conflicting domains in backbone layers, \sumabbrs capture different distributions without causing interference.

\noindent\textbf{Low-precision Friendly Architecture} To enable stable low-precision training (e.g., FP8) while maintaining quality, we implement several architectural modifications:

\begin{itemize}[leftmargin=*,noitemsep,topsep=0pt]
  \item \textbf{Bias-less Backbone Layers} Remove bias terms from linear and normalization layers to prevent unbounded growth and numerical overflow in deep MLPs~\citep{groeneveld2024olmoacceleratingsciencelanguage}.

  \item \textbf{Stabilized Interactions} Apply proper normalization to input/output tensors in modules like Deep Cross Nets~\citep{wang2021dcn} that cause training instability in DHEN ensembles.

  \item \textbf{Swish RMSNorm (SwishRN)} Combine Swish activation~\citep{Ramachandran2017SwishAS} with RMSNorm for FFN layers for  expressivity and stability: $
    X_{out} = RMSNorm(X_{in}) \odot Sigmoid(RMSNorm(X_{in}))$. RMSNorm avoids catastrophic cancellation issues of LayerNorm when elements are near the mean, while the additional RMSNorm on self-gating improves numerical stability. 
\end{itemize}

\subsubsection{\textbf{Supervision Changes}}
\label{sec:supervision-change}

When portfolios are consolidated, models must learn representations that work well across multiple domains simultaneously. Without proper supervision, models tend to optimize each domain's loss independently, leading to conflicting gradients and suboptimal cross-domain knowledge transfer.

\mpartition introduces a cross-domain correlation loss to align representation spaces across consolidated domains: $\mathcal{L}_{corr}(X,Y) = 1 - \frac{Cov(X,Y)}{\sigma_{X}\sigma_{Y}}$, 
where $X$ represents the ground truth label distribution across domains in a batch and $Y$ represents the predicted label distribution from the model. $Cov$ denotes covariance and $\sigma$ denotes standard deviation. This auxiliary loss is computed per batch and added to the primary objective during training.

The correlation loss encourages the model to maintain consistent prediction patterns across domains by: (1) Preventing domain isolation. Without this loss, models can optimize per-domain objectives independently, leading to conflicting representations rather than beneficial knowledge sharing across consolidated portfolios; (2) Balancing domain confidence. The loss prevents models from becoming overconfident in domains with abundant data while remaining underconfident in data-sparse domains, promoting more balanced learning across the consolidated portfolio. By aligning prediction distributions, the shared backbone learns representations that generalize better across all consolidated domains.

When optimizing multiple objectives simultaneously, gradient magnitudes can vary significantly across tasks, causing training instability. We address this using MetaBalance~\cite{he2022metabalance}, which adaptively weighs gradients to ensure stable multi-objective optimization. We use a second-order optimizer~\citep{gupta2018shampoopreconditionedstochastictensor} to optimize \sumabbrs.

\subsection{\textbf{Enabling Inference-Time Knowledge Transfer via \kdap}}
\label{sec:ktap}

Knowledge distillation is commonly used to improve low-latency, user-facing models via powerful teachers. However, traditional soft-label distillation transfers knowledge only during training. 

We design \kdap to address this fundamental limitation through asynchronous precompute and feature-based knowledge transfer, enabling continuous knowledge transfer from teacher models to smaller student models during both training and inference:

\begin{itemize}[leftmargin=*,noitemsep,topsep=0pt]
  \item \textbf{Background Teacher Computation} Teacher \sumabbrs maintain background jobs that continuously evaluate the most relevant items (identified by earlier-stage ranking services) for each user, storing the final backbone layer embeddings for each user-item pair with time-to-live (TTL) metadata.
  
  \item \textbf{Student Query Mechanism} During both training and inference, student \sumabbrs query each user-item pair to retrieve precomputed teacher embeddings, checking TTL validity (typically a few hours) to ensure freshness.
  
  \item \textbf{Embedding Integration} Valid teacher embeddings are directly incorporated as additional input features to the student model. For expired or missing embeddings, the user-item pair is queued for the teacher's next refresh cycle, while zero tensors serve as placeholders in the student.
  
  \item \textbf{Dual Knowledge Transfer} In addition to feature-based transfer, teacher prediction logits are provided to students for traditional label-based distillation, creating a comprehensive knowledge transfer mechanism.
\end{itemize}

Unlike traditional knowledge distillation that relies solely on teacher labels for loss computation, \kdap incorporates both precomputed embeddings as model inputs and teacher labels in the loss function. This dual approach significantly enhances the knowledge transfer ratio during both training and inference, providing richer supervision signals than conventional methods.

\kdap achieves an optimal balance between staleness and computational efficiency by leveraging the natural stability of user interests over short periods. Since user preferences typically remain consistent within a few hours, teacher embeddings retain their relevance, enabling effective student model enhancement during inference without requiring real-time teacher computation.

To ensure cost-effective deployment, \kdap employs caching and clustering techniques that minimize computational overhead. The system provides flexible resource allocation by allowing dynamic adjustment of precomputed embedding volumes based on real-time traffic patterns and business priorities, making it highly scalable across varying workloads.

\kdap incorporates feature clipping~\cite{tao2024featureclippinguncertaintycalibration} on teacher embeddings and label smoothing~\cite{müller2020doeslabelsmoothinghelp} on teacher labels to further improve training stability and convergence properties.

\subsection{Efficiency Optimizations}
\label{sec:efficiency_optimizations}
This section details efficiency improvements for \mthreen.

\subsubsection{\textbf{Distributed Training Optimizations}}
\label{sec:training_opts}
We utilize hybrid parallelism~\citep{neo} with TorchRec~\citep{ivchenko2022torchrec} for embedding table sharding and FSDP ~\citep{fsdp,liang2024torchtitanonestoppytorchnative} for dense parameter synchronization. Small parameters use DDP~\citep{li2020pytorchdistributedexperiencesaccelerating} with pre-allocated static GPU storage to eliminate padding overhead and enable efficient gradient bucketing.

\subsubsection{\textbf{Low-precision Training and Inference}}
\label{sec:low_precision_training_inference}
Leveraging architectural changes from \S\ref{sec:arch-change}, we employ mixed-precision FP8/BF16/FP32 training and FP8 inference. Linear layer GEMMs use FP8 or BF16 based on performance trade-offs, while normalization layers upcast to FP32 for stability. Inference converts weights to FP8 for reduced storage and quantization costs.

Unlike block-wise scaling approaches~\cite{deepseekai2025deepseekv3technicalreport}, we use tensor-wise or row-wise scalers with FBGEMM kernels~\cite{khudia2021fbgemmenablinghighperformancelowprecision}. Fast accumulation~\cite{colfax_deepseek_fp8_2025} is disabled during training but enabled for inference.

\subsubsection{\textbf{Optimized GPU Kernels}}
\label{sec:gpu_kernel}
We combine automatic optimization via Torch Compile with manual kernel fusion for SwishRN operators. Our BlockNorm strategy operates on local GEMM tiles to avoid cross-SM synchronization, fusing with preceding GEMM kernels to leverage L1/L2 cache locality. We replace Sigmoid with HardSwish~\citep{howard2019searchingmobilenetv3} to eliminate expensive exponential operations.

\subsubsection{\textbf{Iterative Execution Refinements with \metasketch}}
\label{sec:metasketch}
To optimize model execution efficiency without compromising quality, we propose \metasketch, a unified search framework maximizing the (model quality, throughput) tuple given latency budget $T$ and quality threshold $Q$.

\noindent\textbf{Alternating Optimization Phase:} \metasketch's search space consists of FSDP sharding strategies and model hyperparameters. It alternates solving phases via beam search of width $K$, with each phase optimizing either hyperparameters or sharding strategies. Each phase contains $S$ steps guided by parallel Bayesian optimizers~\citep{bayesianopt}, with configurations violating $T$ or $Q$ constraints receiving zero scores. The algorithm initializes $K$ seeding configurations, then enters sharding strategy optimization where hyperparameters are fixed and $S$ mutated sharding strategies are evaluated on cluster. Initial seeds are replaced by higher throughput configurations before transitioning to hyperparameter search phase. This alternative optimization strategy allows us to make better use of clusters, such that hyperparameters-related tunings only need to happen locally.

\noindent\textbf{Scaling Law-Guided Search Space Reduction:} To accelerate search, \metasketch leverages established scaling laws~\citep{shin2023scaling,kaplan2020scaling,hoffmann2022trainingcomputeoptimallargelanguage} to shrink hyperparameter spaces. For Wukong, we simultaneously scale output embedding counts in LCB and FMB; for transformers, hyperparameter modifications show minimal quality impact~\cite{narang2021transformer} with a fixed FLOPs budget.

\noindent\textbf{Dynamic Programming Bootstrapping:} We bootstrap FSDP sharding strategy search by profiling execution latency $T_b(s_l)$ and memory usage $R_b(s_l)$ for each layer $l$ with batch size $b$ and sharding strategy $s_l$, plus communication latency $C(s_l)$. Using dynamic programming with GPU memory capacity $R$, we find optimal configurations $ANS_b^*$ by enumerating batch sizes (powers of 2).

\begin{small}
\begin{center}
\textit{Border condition:}
\end{center}
\begin{equation*}
    OPT_b[l,r] = \infty, ANS_b[l,r] = 0, \forall_x x \leq 0 \lor r > R \lor x > L
\end{equation*}
\begin{center}
\textit{Recurrence:}
\end{center}
\begin{equation*}
    s^*_{l+1, r}=argmin_{s \in S} OPT_b[l, r - R_b(s)] + T_b(s) + C(s)
\end{equation*}
\begin{equation*}
    OPT_b[l+1,r] = min_{s \in S} OPT_b[l, r - R_b(s)] + T_b(s) + C(s)
\end{equation*}
\begin{equation*}
    ANS_b[l+1,r] = ANS_b[l, r - R_b(s^*_{l+1,r})] \leftarrow s^*_{l+1,r}
\end{equation*}
\begin{center}
\textit{Output (algorithm complexity $O(|K| \times |S| \times LR)$ given $b$:)}
\end{center}
\begin{equation*}
    ANS_b^*=ANS_b[L-1, argmin_r OPT_b[L-1, r]]
\end{equation*}
\end{small}

\section{Evaluation}
\label{sec:evaluation}
We evaluate each \mthreen component's contribution to cost-efficient recommendation at industry scale.

\subsection{Evaluation Setup}
\label{sec:eval_setup}
\noindent\textbf{Datasets and Baselines} We evaluate \mthreen on one public dataset (Kuaishou~\citep{kuaivideo}) and multiple internal datasets (1-3K features each), comparing against 10 state-of-the-art baselines: AFN+~\citep{afn}, AutoInt+~\citep{song2019autoint}, DLRM~\citep{dlrm}, DCNv2~\citep{wang2021dcn}, FinalMLP~\citep{mao2023finalmlp}, MaskNet~\citep{wang2021masknet}, xDeepFM~\citep{lian2018xdeepfm}, BST~\citep{chen2019behaviorsequencetransformerecommerce}, APG~\citep{NEURIPS2022_9cd0c571}, and Wukong~\citep{zhang2024wukong}.

\noindent\textbf{Metrics} We report AUC and binary cross-entropy (BCE) loss for public datasets, and relative BCE improvement over baselines for internal datasets (0.01\% improvement is significant at our scale). Throughput is measured in queries per second (QPS). Teacher models are applied equally across all internal evaluations, and they have negligible cost (teacher:student GPU ratio $\approx$ 1:100) hence is omitted.

We provide an ablation of overall gain in \S\ref{sec:deployment_ablation}.

\subsection{Effectiveness of Data Consolidation}
\label{sec:eval_data_consolidation}
We evaluate \maw and \pofs on their abilities to consolidate multi-attribution window datasets and select pareto-optimal features using industry-scale datasets.

\begin{figure}[t]
        \centering
        \includegraphics[width=.7\columnwidth]{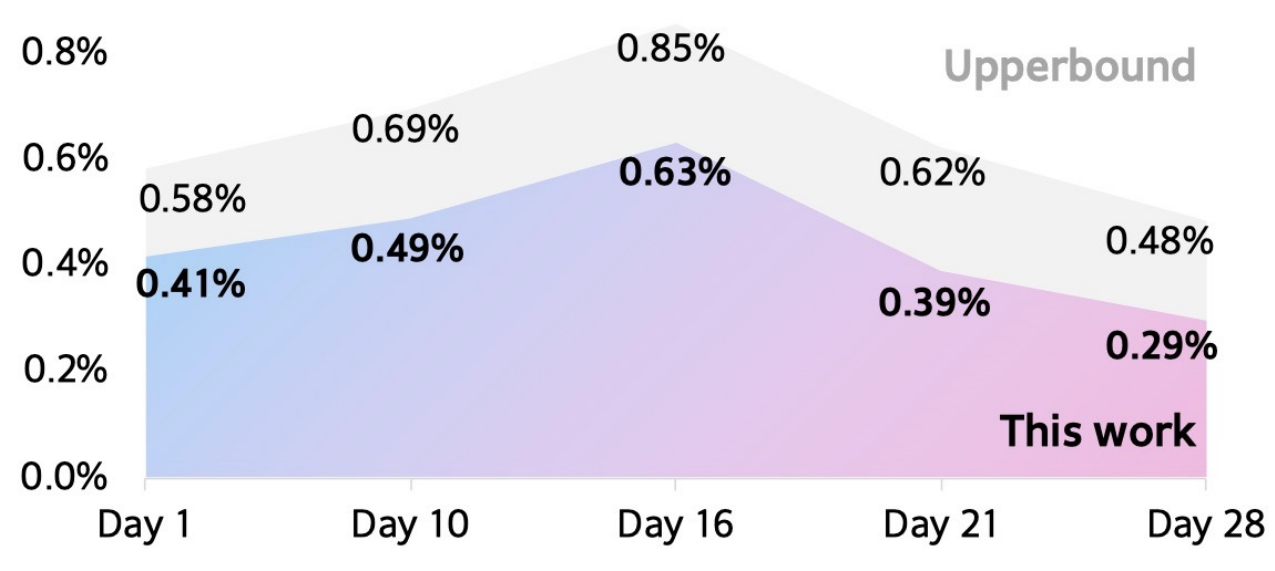}
        \vspace{-1em}
        \caption{Relative Loss Improvement (\%) of \maw across various dates in a month, compared to an upperbound.}
        \label{fig:maw_evaluation}
    \vspace{-1.1em}
\end{figure}

\noindent\textbf{\maw} We apply \maw to consolidate two datasets with 90-minute and 1-day attribution windows, assessing its ability to balance freshness and correctness. 
To establish performance bounds, we simulate an ideal case where all ground-truth labels are immediately available as ``Upperbound''. We evaluate models under daily incremental training over one month. Our baseline uses only the 1-day attribution window model (we omit the 90-minute-only model for visual clarity as it consistently shows 1\% regression).
We compare \maw against existing solutions: delayed feedback modeling~\citep{wang2020delayed,chen2022asymptotically}, continuous training with negative samples~\citep{ktena2019addressing,wang2020delayed}, and label correction with reweighing~\citep{chen2022asymptotically}. All baseline approaches requiring multi-pass training exhibited severe divergence or loss regression on our datasets while being less efficient. \Cref{fig:maw_evaluation} shows \maw achieves consistent improvements across all evaluation dates. It's worth noting that the effectiveness of \maw extends beyond the two-window consolidation. Specifically, we see additional gain on the 7d head when consolidating 90min/1d/7d combination. Adding more consolidation windows to the mix yield additional yet diminishing returns.

\noindent\textbf{\pofs} We apply \pofs to select approximately 2K features from a pool of 12K features in a consolidated dataset.

We compare against a standard weighted loss-based feature selection baseline~\citep{debnath2008feature,jain2016extreme}. \Cref{tab:pofs_table} shows relative loss improvements across 10 models spanning 4 optimization goals: \pofs generalizes well across diverse tasks and combinations compared to traditional approaches.
    
\begin{table}[t]
    \centering
    \footnotesize
    \begin{center}
        \begin{tabular}{c c}
        \toprule
         \textbf{Task Type}  & \textbf{Loss Improvements (\%)}  \\
            \midrule
            CTR of multiple domains & 0.2 $\sim$ 0.5 \\
            CVR of multiple domains & 0.12$\sim$0.17 \\
            CTR+CVR of a single domain & 0.1 $\sim$ 0.5  \\
            CTR+Quality of a single domain & 0.06  \\
            \bottomrule
        \end{tabular}
    \end{center}
    \caption{Evaluation of \pofs.}
    \vspace{-3.1em}    
    \label{tab:pofs_table}    
\end{table}

        

\subsection{Effectiveness of \sumabbrs}
We focus our evaluation on the quality gains from adopting \sumabbrs on a public dataset and industry-scale datasets.

\noindent\textbf{Open Source \mdmo Dataset: KuaiVideo} We simultaneously predict the labels for $like$, $follow$, and $click$ to test the models' \mdmo performance, following established approaches~\citep{bars, bars2}.
We report final test performance in \Cref{tab:mdmo_kuaivideo} after learning rate tuning for all models. Best and second-best performing models are bolded and underlined. \sumabbrs match or outperform state-of-the-art performance in 7 out of 8 metrics with comparable or lower complexity.

\begin{table*}[!t]
\footnotesize
\resizebox{.85\textwidth}{!}
{
    \begin{tabular}{c c c c c c c c c c c}
        \toprule
         \multirow{2}{*}{\textbf{Model}} & \multicolumn{4}{c}{\textbf{AUC}} & \multicolumn{4}{c}{\textbf{Loss}} & \multicolumn{2}{c}{\textbf{Complexity}} \\        
                                         & Click  & Follow & Like   & \underline{AVG}    & Click  & Follow & Like   & \underline{AVG}    & MFLOPs & MParams \\ 
        \midrule
        AFN+                             & 0.7172             & 0.7703 & 0.8466 & 0.7780 & 0.4572          & 0.0074 & 0.8466          & 0.1604 & 10.96 & 79.60 \\    
        AutoInt+                         & 0.7181             & 0.7882 & 0.8725 & 0.7929 & 0.4607          & 0.0075 & 0.0156          & 0.1617 & 79.27 & 41.75 \\
        DLRM                             & 0.7088             & 0.6743 & 0.7734 & 0.7188 & 0.4874          & 0.0081 & 0.0175          & 0.1710 & 1.996 & 39.24 \\
        DCNv2                            & 0.7225             & 0.7954 & 0.8804 & 0.7995 & 0.4534          & \textbf{0.0073} & \underline{0.0152} & 0.1586 & 9.159 & 40.44 \\
        FinalMLP                         & 0.7176             & 0.7627 & 0.8624 & 0.7809 & 0.4690          & 0.0080 & 0.0163          & 0.1645 & 12.22 & 571.4 \\
        MaskNet                          & 0.7133             & 0.7143 & 0.8599 & 0.7625 & 0.4650          & 0.0077 & 0.0156          & 0.1627 & 4.299 & 39.63 \\
        xDeepFM                          & 0.7189             & 0.7704 & 0.8706 & 0.7866 & 0.4642          & 0.0079 & 0.0156          & 0.1626 & 6.810 & 51.60 \\
        APG (DeepFM)                     & 0.7066             & 0.7464 & 0.8515 & 0.7682 & 0.4915          & 0.0080 & 0.0166          & 0.1720 & 11.74 & 52.40 \\
        BST                              & 0.7217             & 0.7664 & 0.8707 & 0.7863 & \textbf{0.4512} & 0.0076 & 0.0153          & \underline{0.1581} & 326.1 & 40.63 \\
        Wukong                           & \underline{0.7251} & 0.7947 & \underline{0.8842} & 0.8014 & 0.4580          & 0.0075 & 0.0155          & 0.1603 & 22.62 & 42.59  \\
        \hline
        \textbf{\sumabbr} (Expert Tuned)          & \textbf{0.7281}    & \textbf{0.7997} & 0.8793 & \underline{0.8024} & \underline{0.4513} & \textbf{0.0073} & 0.0154 & \textbf{0.1580} & 25.54 & 43.08 \\
        \textbf{\sumabbr} (\metasketch)           & 0.7249 & 0.7984 & \textbf{0.8861} & \textbf{0.8031} & 0.4529 & \textbf{0.0073} & \textbf{0.0151} & 0.1584 & \textbf{1.575} & \textbf{39.17} \\ 
        \bottomrule
    \end{tabular}
}
    \caption{Test performance on KuaiVideo across 3 tasks. \sumabbrs achieve best performance with least resources. }
    \vspace{-2em}
    \label{tab:mdmo_kuaivideo}

\end{table*}

\begin{figure}[t]
    \begin{center}
    \includegraphics[width=\columnwidth]{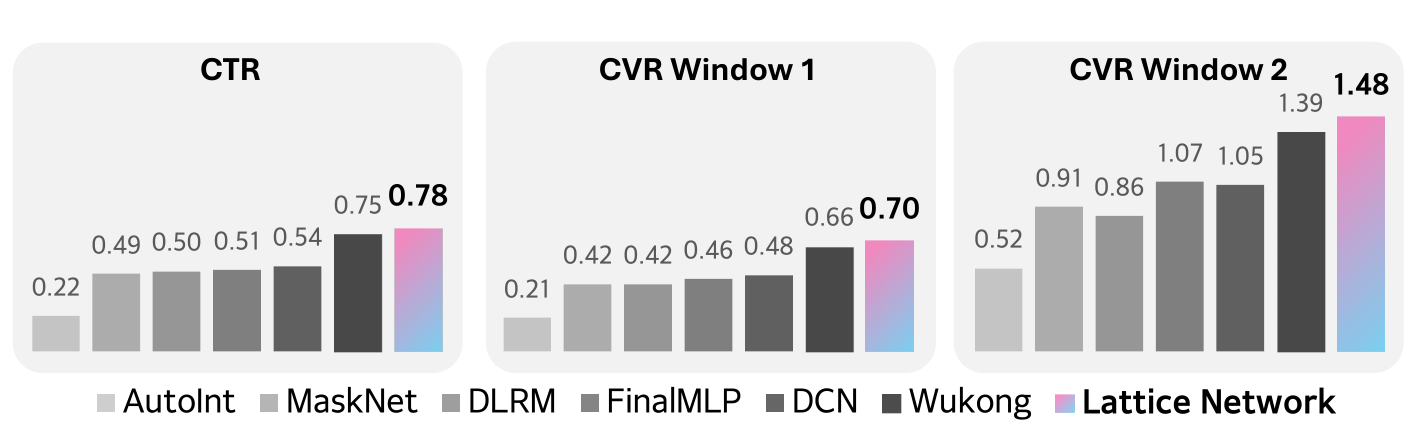}
    \vspace{-2em}    
    \caption{Relative Loss Improvement (\%) over AFN on an industry-scale dataset.}
    \vspace{-1em}    
    \label{fig:nonseq_ne}  
    \end{center}    
\end{figure}

\begin{table}[t]
    \footnotesize
    \setlength{\tabcolsep}{0.8em}
    \centering
    \begin{tabular}{c c c c c c c c c c}
        \toprule
         \textbf{CTR} & \textbf{CONV} & \multicolumn{2}{c}{\textbf{QLT}} & \multicolumn{2}{c}{\textbf{LG}} & \multicolumn{4}{c}{\textbf{PE}} \\
                      &              & T1 & T2 & T1 & T2 & T1 & T2 & T3 & T4 \\
        
        \midrule
        0.27 & 0.36 & 0.38 & 0.39 & 0.48 & 0.63 & 0.59 & 0.68 & 1 & 1.14 \\
        
        \bottomrule
    \end{tabular}
    \caption{Relative Loss Improvement (\%) of \sumabbr over Wukong on an industry-scale, mixed-sequence dataset.}
    \vspace{-2em}
    \label{tab:seq_ne}    
\end{table}



\noindent\textbf{Industry-Scale Dataset: Non-sequence Data} We upscale selected models to 30-40~GFLOPs/sample and evaluate on our industry-scale dataset with 100B samples for scalability testing. We use three tasks: one CTR and two CVR tasks.
We present results in \Cref{fig:nonseq_ne}, using AFN as a baseline. \sumabbrs continue to significantly outperform other state-of-the-art models, improving loss by up to 1.14\%, demonstrating the scalability of \sumabbrs and the quality improvements from the architectural changes described in \S\ref{sec:arch-change}.

\noindent\textbf{Industry-Scale Dataset: Mixed Sequence and Non-sequence Data} We compare \sumabbr's performance with scaled-up Wukong on a mixed-format dataset with 50B samples merged from domains with mixed sequence data. We predict O(10) objectives among click (CTR), conversion (CONV), and ads quality (QLT) types. \Cref{tab:seq_ne} summarizes results for overall CTR, CVR, QLT, and selectively reports two conversion types: lead generation (LG) and post engagements (PE) to demonstrate effectiveness on finer-grain breakdowns. \sumabbrs significantly outperform Wukong in every task with comparable complexity, demonstrating the effectiveness of interleaved learning (\S\ref{sec:network_backbone}) and our architectural changes (\S\ref{sec:arch-change}).

\subsection{Effectiveness of Portfolio Consolidation}
We evaluate the effectiveness of portfolio consolidation, guided by \mpartition, in a challenging scenario involving CTR prediction on two product surfaces (A and B). These were originally handled by two separate models with 1.33T and 0.71T parameters and training complexity of 20~GFLOPs and 12~GFLOPs/sample, respectively.
We merge the task modules, data, and features as necessary for consolidation, building on the 20~GFLOPs baseline. \Cref{tab:ig_fb_omnifm} shows an immediate benefit for Domain B at the cost of quality loss for Domain A -- a typical manifestation of cross-domain interference in multi-objective setups. However, applying supervision changes (\S\ref{sec:supervision-change}) produces a significant performance boost that surpasses the unconsolidated baseline.
Overall, portfolio consolidation guided by \mpartition achieves $1.5\times$ FLOPs and $1.04\times$ parameter savings\footnote{Parameter count is dominated by embedding table size, determined by the total number of features consolidated, hence smaller savings compared to FLOPs.} with significant performance improvements in both domains.
The minimal complexity increase indicates that \mthreen portfolio consolidation facilitates knowledge transfer between domains, underscoring its key advantage: \textit{the cost of serving a consolidated portfolio is significantly lower than serving portfolios separately}.

\begin{table}[t]
    \footnotesize
    \setlength{\tabcolsep}{1.2666em}
    \centering
    \begin{tabular}{c c c c c}
        \toprule
         \multirow{2}{*}{Consolidation Type} & \multicolumn{2}{c}{\textbf{Loss Improvements}} & \textbf{FLOPs} & \textbf{Params} \\
                   & CTR on A & CTR on B & Saving & Saving \\
        
        \midrule
        Data & -0.15\% & 0.36 \% & $1.59\times$ & $1.04\times$ \\
        Data + Aux Loss & 0.13\% & 0.46 \% & $1.50\times$ & $1.04\times$ \\
        \bottomrule
    \end{tabular}
    \caption{Effectiveness of Portfolio Consolidation.}
    \vspace{-2em}
    \label{tab:ig_fb_omnifm}    
\end{table}

\subsection{Inference-time Knowledge Transfer Gains}
We measure improvements using \kdap, by comparing the loss improvements of a distilled \sumabbr over an non-distilled version via: (1) traditional soft-label distillation; (2) \kdap with a typical observed cache hit rate of 0.6 and (3) \kdap with an ideal case of 1.0 cache hit rate, on a CTR task. We set TTL to 6 hours. The teacher, student and baseline models are \sumabbrs with complexity of 2.7~GFLOPs and 0.42~GFLOPs. 

Shown in Table~\ref{tab:kdap}, we observe over a $1.3\times$ boost in knowledge transfer efficiency in typical settings, harnessing most of the benefit in the ideal case of 100\% hit rate. 

\begin{table}[t]
    \centering
    \footnotesize
    \begin{center}
        \begin{tabular}{c c}
        \toprule
         \textbf{Knowledge Transfer Type}  & \textbf{Loss Improvements (\%)}  \\
            \midrule
            Soft-label Distillation & 0.12 \\
            \kdap (Hit Rate 0.6) & 0.24 \\
            \kdap (Hit Rate 1.0, Upperbound) & 0.43 \\
            \bottomrule
        \end{tabular}
    \end{center}
    \caption{Effectiveness of \kdap.}
    \vspace{-3em}    
    \label{tab:kdap}    
\end{table}


 \begin{figure}
    \centering
    \includegraphics[width=.75\columnwidth]{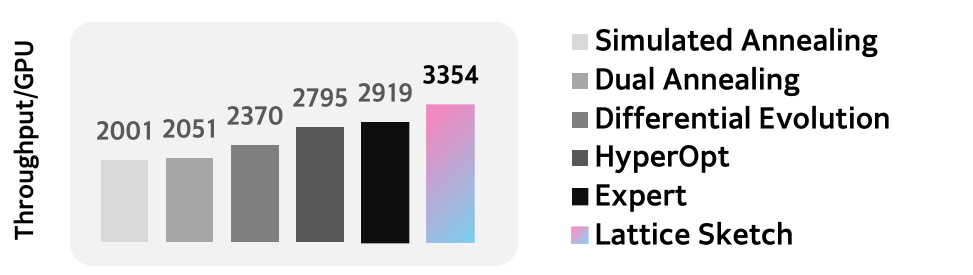}
    \vspace{-1em}            
    \caption{\metasketch improves throughput by 20\% on 128 A100 GPUs through iterative refinements.} 
    \label{fig:osdp}
    \vspace{-2em}    
\end{figure}

\subsection{Cost-efficiency Improvements}
\label{sec:eval_cost_efficiency}
We evaluate \metasketch's efficacy on hyperparameter tuning and throughput improvements.

\noindent\textbf{\metasketch on Quality} We set up a search space for \metasketch with 4 trillion choices and apply it to \sumabbr on the KuaiVideo dataset. We set the objective to $\max O_{VM} = \max \frac{AUC}{f^{0.003}}$ on the validation set, reflecting our expectation that \mthreen's AUC scales linearly with $f^{0.003}$. This modification  biases the search slightly toward lower-latency configurations that outperform the scaling baseline in efficiency (as prescribed by the $f^0.003$ scaling law slope). We use \metasketch with a budget of 1200 steps. We present results in \Cref{tab:mdmo_kuaivideo}. The tuned model achieves state-of-the-art results in 4 out of 8 metrics, outperforming the expert-tuned model with $17\times$ fewer FLOPs.

\noindent\textbf{\metasketch on Training Throughput} We evaluate \metasketch's throughput improvement by simultaneously adjusting batch size and FSDP parallelization strategy for each DWFB on 128 Nvidia A100 GPUs for an 8-layer \sumabbr.
\Cref{fig:osdp} summarizes the results. Compared to an expert-tuned baseline and 4 state-of-the-art optimizers---including HyperOpt~\citep{bergstra2013hyperopt}, annealing~\citep{kirkpatrick1983optimization}, dual annealing~\citep{scipyDual_annealingx2014}, and differential evolution~\citep{storn1997differential}---given the same search budget of 500 attempts, \metasketch delivers up to $1.6\times$ better throughput.

\noindent\textbf{Other Efficiency Improvements} When coupled with the optimizations in \S\ref{sec:training_opts}, \sumabbr achieves an additional 26\% faster inference speed with FP8 inference, where 10\% is contributed by kernel optimizations.

\section{Industrial Deployment and Impacts}
\label{sec:deployments}
The surging demand for recommendation systems in recent years~\citep{zhai2024actions, wu2022sustainable, neo, DBLP:journals/corr/abs-1811-09886} has made \mthreen essential for sustained growth of recommendation at Meta.

\subsection{Deployment Baseline and Methodology}
\label{sec:deployment_methodology}
We deployed \mthreen across a critical set of ads model types serving global Meta users across platforms. Our baseline is the prior production system without \mthreen, where each portfolio was served by a separate model trained on isolated datasets. During testing, both \mthreen and the baseline predict CTR/CVR metrics for live traffic in the form of (user, ad) pairs, with the highest-ranked pairs presented to users based on predictions and other factors. 

We measure \mthreen's impact using: (1) revenue-driving top-line metrics, (2) relevance metrics such as (CTR, CVR, ads quality (user satisfaction~\cite{adsquality}), and (3) cost metrics (capacity savings in power usage). Our blog posts~\cite{meta_lattice_2024, meta_andromeda_2024} have more details.

\subsection{Results and Ablation}
\label{sec:deployment_ablation}
\mthreen has delivered substantial real-world impact: 10\% top-line metric gains, 11.5\% improvement in ads quality, and up to 6\% increase in ad conversions while achieving 20\% capacity savings.

We ablated the top-line metric contributions as follows: \mpartition (36\%), \maw (11\%), \pofs (13\%), \sumabbrs (23\%), and \kdap (17\%). The capacity savings result entirely from reducing model count and sharing compute through improved efficiency across portfolios, and has factored in the cost associated with tuning \mthreen.

\subsection{Deployment Trade-offs}
\label{sec:deployment_trade_offs}
Our deployment experience demonstrates that \mthreen achieves cost reduction without quality trade-offs. Small models benefit from upscaling, while potential conflicts between objectives and domains are prevented by \mpartition (\S\ref{sec:partitioner}) and mitigated through auxiliary loss (\S\ref{sec:supervision-change}), parameter untying (\S\ref{sec:arch-change}), and domain-specific towers (\S\ref{sec:tower_arch}). To balance model complexity and latency, \mthreen employs efficient architectures (\S\ref{sec:arch-change}), low-precision (BF16/FP8) execution, optimized GPU kernels (\S\ref{sec:low_precision_training_inference}-\ref{sec:gpu_kernel}), and optimal execution strategies via \metasketch (\S\ref{sec:metasketch}).

\section{Discussion and Limitations}
\label{sec:discussion}
\textbf{Comparison with State-of-the-Art} The holistic approach to portfolio, data, and model consolidation distinguishes \mthreen from existing methods that focus solely on model design~\cite{zhang2024m3oe,dredze2010multi,misra2016cross,ma2018modeling,li2019multi} or efficiency~\cite{dmt, neuroshard, ivchenko2022torchrec, neo} in isolation, addressing a critical gap in the literature regarding data curation for \mdmo learning in industrial settings and improving upon feature selection schemes that directly compute importance scores (e.g., using Shapley values~\citep{roth1988shapley}) and selecting top-K features with lowest loss~\citep{ma2018entire,xi2021modeling,yasuda2022sequential} to avoid loss-gaming. Compared to neural architecture search (NAS)~\cite{elsken2019neural}, \metasketch leverages established scaling laws and does not require waiting for slow quality signals, enabling faster iteration.

\noindent\textbf{Limitations} While we strive to share generalizable deployment insights with the broader community, some findings contain Meta-specific elements. For example, Meta's business requirements and partner relationships define the problem spaces of by \mthreen.
\section{Conclusion}
Through comprehensive model space redesign that integrates portfolio consolidation, data unification, architectural innovations, and efficiency optimizations, Meta \mthreen demonstrates substantial real-world impact: 10\% improvement in revenue-driving top-line metrics, 11.5\% enhancement in ads quality, 6\% increase in conversion rates, and 20\% capacity savings. These results validate \mthreen's effectiveness in bridging the gap between theoretical advances in recommendation systems and practical industry deployment.
\end{sloppypar}

\bibliographystyle{ACM-Reference-Format}
\balance
\bibliography{sample-base}

\appendix
\section{Appendix}

\subsection{Detailed Experimental Setup}
\subsubsection{Datasets}
\label{appendix:setup}
We evaluate \mthreen on both public and internal datasets. The public dataset KuaiVideo is an representative competition dataset released by Kuaishou~\cite{kuaivideo} as used in ~\cite{zhang2024wukong, 10.1145/3343031.3350950, 10.1145/3580305.3599785}. This dataset has 13M entries with 8 features. We use this dataset as a standard benchmark for recent state-of-the-art models in multi-objective recommendation. We use the train/test split provided by the BARS~\cite{10.1145/3477495.3531723} benchmark suite, and we further perform 9 to 1 train and validation split. We use and extend the FuxiCTR framework~\cite{10.1145/3459637.3482486} for experimentation on public dataset. We use two internal datasets, with and without sequence features to evaluate model performance. The dataset without event feature has about 1K features. The dataset with event features contains roughly 2K features and 9 event sources. For evaluating data strategy, we use an internal dataset with about 2K nonsequence features, selected from a pool of 12K features.

\subsubsection{Metrics and Objectives}
For KuaiVideo, we report AUC and loss; for internal datasets, we report improved normalized entropy (NE)~\cite{10.1145/2648584.2648589} over a baseline, following prior arts. 

Due to lack of a common fundamental task across all recommendation tasks, we use a collection of representative tasks to \textit{approximate} a foundational task from which all downstream tasks can benefit. For the Kuaishou dataset, we predict three tasks: \textit{is\_like}, \textit{is\_follow} and \textit{is\_click} and use the average loss as the final loss. For internal dataset, we create 3 tasks derived from a combination of  click and conversion rate prediction across various attribution windows, and we train each model sufficiently long enough to draw conclusions and report metrics.

\subsubsection{Baselines}
We focus on comparing with recent state-of-the-art recommendation models including AFN+ \cite{afn}, AutoInt+ \cite{song2019autoint}, DLRM \cite{dlrm}, DCNv2 \cite{wang2021dcn},
FinalMLP \cite{mao2023finalmlp}, MaskNet \cite{wang2021masknet}, xDeepFM \cite{lian2018xdeepfm}, BST~\cite{chen2019behaviorsequencetransformerecommerce}, APG~\cite{NEURIPS2022_9cd0c571} and Wukong~\cite{zhang2024wukong}. For public dataset, we use the best-tuned config from BARS if available, otherwise we use the configuration used by ~\cite{zhang2024wukong}. For internal dataset, we use the model tunings adopted by ~\cite{zhang2024wukong}. We report model complexity and parameter count for fair comparison. Note that we may use different model configurations to highlight the effectiveness of different components, which we detail in their respective sections.

\end{document}